\documentclass{article}
\usepackage{ijcai26}

\usepackage{times}
\usepackage{soul}
\usepackage{url}
\usepackage[hidelinks]{hyperref}
\usepackage[utf8]{inputenc}
\usepackage[small]{caption}
\usepackage{graphicx}
\usepackage{amsmath}
\usepackage{amsthm}
\usepackage{booktabs}
\usepackage{algorithm}
\usepackage{algorithmic}
\usepackage[switch]{lineno}
\usepackage{xcolor}
\usepackage{amssymb}

\usepackage{tikz}
\usetikzlibrary{positioning,calc,fit,backgrounds,arrows.meta,decorations.pathreplacing}

\definecolor{sxzblue}{RGB}{0, 51, 153}

\title{When Vision Meets Graphs: A Survey on Graph Reasoning and Learning}

\author{
Xinjian Zhao$^1$
\and
Wei Pang$^1$\and
Zhixuan Yu$^1$\and
Xiangru Jian$^2$\and
Xiaozhuang Song$^1$\and
Yaoyao Xu$^1$\and 
Zhongkai Xue$^1$\and
Dingshuo Chen$^3$\and
Shu Wu$^3$\and
Philip Torr$^4$\And
Tianshu Yu$^1$\\
\affiliations
$^1$The Chinese University of Hong Kong, Shenzhen \\
$^2$University of Waterloo \\
$^3$Institute of Automation, Chinese Academy of Sciences \\
$^4$University of Oxford \\
\emails
\{xinjianzhao1, weipang, zhixuanyu\}@link.cuhk.edu.cn,
x2jian@uwaterloo.ca, \{xiaozhuangsong1, yaoyaoxu, zhongkaixue\}@link.cuhk.edu.cn,
\{dingshuo.chen, shu.wu\}@nlpr.ia.ac.cn, \\ philip.torr@eng.ox.ac.uk, yutianshu@cuhk.edu.cn
}

\begin{document}

\maketitle

\begin{abstract}
Graphs are a fundamental data structure underlying many problems in the natural and social sciences. Over the past decade, Graph Neural Networks (GNNs) have dominated graph machine learning, supported by solid theoretical foundations. Yet scientists often understand graph structure through vision: chemists read molecular diagrams and social scientists inspect network visualizations. Despite decades of work on graph visualization, most graph learning pipelines still treat graphs purely as symbolic structures, rarely leveraging the visual form of graphs. We argue that this gap deserves renewed attention in the era of powerful vision and vision-language models.
This survey provides a first systematic overview of the emerging area we term \textbf{vision meets graphs}, which treats visual depictions of graphs as first-class inputs for reasoning and learning. We organize existing work into three threads. \emph{Vision for Graph Reasoning} studies how models can use visual depictions of graphs to understand structure and carry out multi-step reasoning. \emph{Vision for Graph Learning} explores how visual features can complement or augment graph encoders beyond known limitations of message passing. \emph{Scientific Graphs} examines domains where standardized depiction conventions support both reasoning and learning. Our goal is to clarify what current methods can and cannot do, and to outline a path toward foundation models that perceive and reason about graphs as scientists do.
\end{abstract}

\section{Introduction}
Graphs are a fundamental data structure in science and engineering~\cite{stoll2025graphbench}. They describe how atoms form molecules, how people form communities, how components form circuits, and how data entities form relational graphs~\cite{wu2020comprehensive,jian-wang-2023-invgc,el2025graph,liu2026deepfake}. A central question across these domains is how to understand and reason about graph structure.

Human scientists have long relied on vision to answer this question. Chemists read molecular diagrams, social scientists inspect network visualizations, and engineers reason with circuit graphs. These visual conventions encode accumulated knowledge about which spatial arrangements make structural patterns easy to perceive. Notably, tasks that are theoretically hard for GNNs, such as testing biconnectivity~\cite{zhang2023rethinking} or distinguishing certain non-isomorphic graphs~\cite{xu2018powerful,horn2021topological,wang2023empirical}, often become almost trivial once a graph is visualized appropriately. Decades of graph visualization research have formalized these layout principles~\cite{tamassia2013handbook}, and scientific disciplines have developed domain-specific depiction standards tightly integrated into expert workflows.

Graph machine learning, however, has largely bypassed this visual tradition. GNNs operate directly on adjacency structure, learning representations through message passing~\cite{khoshraftar2024survey}. LLMs serialize graphs into text to tackle multi-step reasoning~\cite{wang2023can,luo2024graphinstruct,chen2024graphwiz,xu2025graphomni}. Both approaches have driven substantial progress, yet neither sees the visual form of graphs that scientists see. We argue that this gap deserves renewed attention. Recent work suggests that visual representations encode structural information qualitatively different from adjacency-based encodings. Vision encoders pretrained only on natural images can outperform GNNs on tasks requiring global structure understanding~\cite{zhao2025underappreciated}.
The rise of vision-language models (VLMs) has prompted researchers to revisit this visual channel~\cite{zhang2024vision}. Work has emerged on benchmarks for reasoning over graph images~\cite{wei2024gita,li2024visiongraph,zhu2025benchmarking}, methods that incorporate visual features into graph representations~\cite{wei2025open}, and scientific applications leveraging domain-specific depiction conventions~\cite{li2025chemvlm,zhao2025tinychemvl}. This growing body of research spans multiple communities but has developed largely in isolation.

\begin{figure*}[t]
\centering
\resizebox{0.98\textwidth}{!}{%
\begin{tikzpicture}[
    rootfill/.style={fill={rgb,255:red,51;green,65;blue,85}},
    rootborder/.style={draw={rgb,255:red,51;green,65;blue,85}},
    blue1/.style={fill={rgb,255:red,191;green,210;blue,235}},
    blue1b/.style={draw={rgb,255:red,120;green,150;blue,190}},
    blue2/.style={fill={rgb,255:red,215;green,228;blue,245}},
    blue2b/.style={draw={rgb,255:red,150;green,175;blue,210}},
    purple1/.style={fill={rgb,255:red,212;green,200;blue,230}},
    purple1b/.style={draw={rgb,255:red,160;green,140;blue,190}},
    orange1/.style={fill={rgb,255:red,245;green,215;blue,195}},
    orange1b/.style={draw={rgb,255:red,210;green,160;blue,130}},
    blacktext/.style={text=black},
    root/.style={rectangle, rounded corners=5pt, rootfill, rootborder, line width=1pt,
        text=white, font=\sffamily\bfseries\fontsize{9.5}{11}\selectfont,
        minimum width=2.3cm, minimum height=1.05cm, align=center},
    L1/.style={rectangle, rounded corners=4pt, line width=0.7pt, blacktext,
        font=\sffamily\bfseries\fontsize{7.5}{9}\selectfont,
        minimum width=2.4cm, minimum height=0.75cm, align=center},
    L1a/.style={L1, blue1, blue1b},
    L1b/.style={L1, purple1, purple1b},
    L1c/.style={L1, orange1, orange1b},
    L2/.style={rectangle, rounded corners=3pt, line width=0.6pt, blacktext,
        font=\sffamily\fontsize{7}{8.5}\selectfont,
        minimum width=2.2cm, minimum height=0.65cm, align=center},
    L2a/.style={L2, blue2, blue2b},
    L2b/.style={L2, purple1, purple1b},
    L2c/.style={L2, orange1, orange1b},
    L3/.style={rectangle, rounded corners=2.5pt, line width=0.5pt, blacktext,
        font=\sffamily\fontsize{6.5}{8}\selectfont,
        minimum width=2cm, minimum height=0.55cm, align=center},
    L3a/.style={L3, blue2, blue2b},
    leafWithL3/.style={rectangle, rounded corners=2.5pt, line width=0.5pt, blacktext,
        font=\sffamily\fontsize{6}{7.5}\selectfont,
        align=left, inner xsep=5pt, inner ysep=3.5pt,
        text width=6.8cm, minimum width=7.2cm},
    leafNoL3/.style={rectangle, rounded corners=2.5pt, line width=0.5pt, blacktext,
        font=\sffamily\fontsize{6}{7.5}\selectfont,
        align=left, inner xsep=5pt, inner ysep=3.5pt,
        text width=9.0cm, minimum width=9.4cm},
    leafA/.style={leafWithL3, blue2, blue2b},
    leafAlong/.style={leafNoL3, blue2, blue2b},
    leafB/.style={leafNoL3, purple1, purple1b},
    leafC/.style={leafNoL3, orange1, orange1b},
    conn/.style={draw={rgb,255:red,100;green,110;blue,130}, line width=0.8pt}
]
\def\xR{0} 
\def\xA{3.2} 
\def\xB{6.2} 
\def\xC{9.2}       
\def\xLeafL3{10.4} 
\def\xLeafNoL3{8.2} 
\def\xRight{17.6}  

\node[root] (root) at (\xR, 0) {When Vision\\Meets Graphs};

\node[L1a] (s1) at (\xA, 2.0) {Vision for\\Graph Reasoning};
\node[L2a] (gt) at (\xB, 2.65) {Graph-Theoretic\\Tasks ({\color{blue}\S\ref{SEC:VGR1}})};
\node[L3a] (eval) at (\xC, 3.25) {Benchmarks  \&\\Methods};
\node[L3a] (robust) at (\xC, 2.20) {Robustness \\ Probes};
\node[leafA, anchor=west] (r1) at (\xLeafL3, 3.25) 
    {GraphTMI~\cite{das2024modality}; VisionGraph~\cite{li2024visiongraph}; GITA~\cite{wei2024gita}; VGCure~\cite{zhu2025benchmarking};GraphVerse~\cite{sun2026graphverse}};
\node[leafA, anchor=west] (r2) at (\xLeafL3, 2.20) 
    {VisGraphVar~\cite{sartori2024visgraphvar}; VGA~\cite{babaiee2025visual}; VNA~\cite{williams2024multimodal}};

\node[L2a] (rw) at (\xB, 1.1) {Real-World\\Graphs ({\color{blue}\S\ref{SEC:VGR2}})};
\node[leafAlong, anchor=west] (r3) at (\xLeafNoL3, 1.1) 
    {\cite{ai2024advancement}; GraphTMI~\cite{das2024modality};  GITA~\cite{wei2024gita}; VNA~\cite{williams2024multimodal}; \cite{white2025using}; Graph-to-vision~\cite{li2025graph}; FlowGen~\cite{shi2026flowgen}}; %

\node[L1b] (s2) at (\xA, -0.4) {Vision for\\Graph Learning};
\node[L2b] (vgnn) at (\xB, 0) {GNN-Centric ({\color{blue}\S\ref{sec:vgl}})};
\node[L2b] (venc) at (\xB, -0.9) {Vision-Centric ({\color{blue}\S\ref{sec:vgl}})};
\node[leafB, anchor=west] (r4) at (\xLeafNoL3, 0) 
    {DEL~\cite{zhao2025graph}; GVN~\cite{wei2025open}};
\node[leafB, anchor=west] (r5) at (\xLeafNoL3, -0.9) 
    {GraphAbstract~\cite{zhao2025underappreciated}};

\node[L1c] (s3) at (\xA, -2.4) {Scientific\\Graphs};
\node[L2c] (scir) at (\xB, -2.0) {Visual\\Reasoning ({\color{blue}\S\ref{sec:sci_reasoning}})};
\node[L2c] (scil) at (\xB, -3.0) {Visual\\Learning ({\color{blue}\S\ref{sec:sci_learning}})};
\node[leafC, anchor=west] (r6) at (\xLeafNoL3, -2.0) 
    {GIT-Mol~\cite{liu2024git}; ChemVLM~\cite{li2025chemvlm}; Mol2Lang-VLM~\cite{tran2024mol2lang}; ChemDFM-X~\cite{zhao2024chemdfm}; TinyChemVL~\cite{zhao2025tinychemvl}; Geneverse~\cite{liu2024geneverse}; LiveProteinBench~\cite{rong2025liveproteinbench}; MicroscopyGPT~\cite{choudhary2025microscopygpt}};
\node[leafC, anchor=west] (r7) at (\xLeafNoL3, -3.0) 
    {Chemception~\cite{goh2017chemception};ImageMol~\cite{zeng2022accurate}; MolEmb~\cite{zhao2026molemb}; DDVR-DDI~\cite{xie2025predicting}; S2VM~\cite{ma2025selfsupervised}; SPROF~\cite{chen2019improve}};

\draw[conn] (root.east) -- ++(0.3,0) |- (s1.west);
\draw[conn] (root.east) -- ++(0.3,0) -- (s2.west);
\draw[conn] (root.east) -- ++(0.3,0) |- (s3.west);
\draw[conn] (s1.east) -- ++(0.25,0) |- (gt.west);
\draw[conn] (s1.east) -- ++(0.25,0) |- (rw.west);
\draw[conn] (gt.east) -- ++(0.25,0) |- (eval.west);
\draw[conn] (gt.east) -- ++(0.25,0) |- (robust.west);
\draw[conn] (eval.east) -- (r1.west);
\draw[conn] (robust.east) -- (r2.west);
\draw[conn] (rw.east) -- (r3.west);
\draw[conn] (s2.east) -- ++(0.25,0) |- (vgnn.west);
\draw[conn] (s2.east) -- ++(0.25,0) |- (venc.west);
\draw[conn] (vgnn.east) -- (r4.west);
\draw[conn] (venc.east) -- (r5.west);
\draw[conn] (s3.east) -- ++(0.25,0) |- (scir.west);
\draw[conn] (s3.east) -- ++(0.25,0) |- (scil.west);
\draw[conn] (scir.east) -- (r6.west);
\draw[conn] (scil.east) -- (r7.west);
\end{tikzpicture}
}%
\caption{Taxonomy of \textit{When Vision Meets Graphs}. We organize the literature into three threads: graph reasoning, graph learning, and scientific graphs. Representative works are listed for each category.}
\label{fig:taxonomy}
\end{figure*}
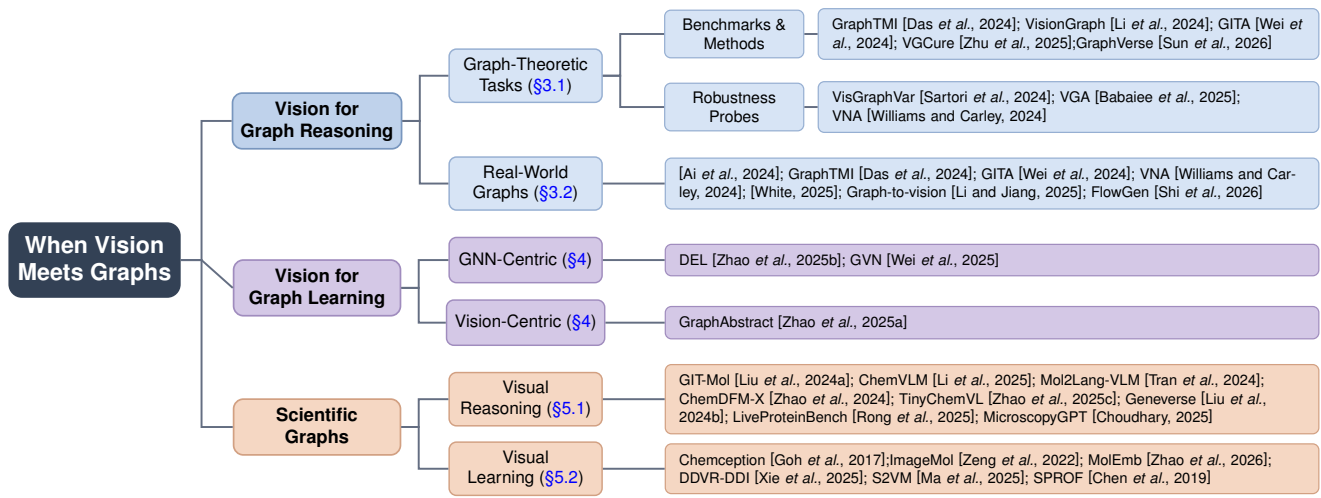

\begin{figure*}[t]
    \centering
    \includegraphics[width=1\linewidth]{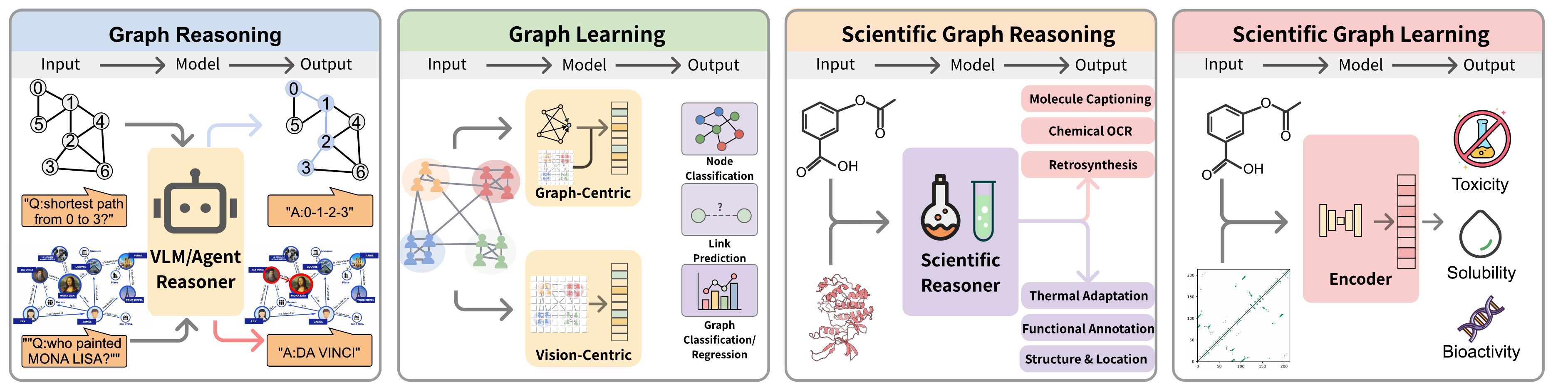}
\caption{Illustrative pipelines for each paradigm. 
\textbf{Graph Reasoning} (\S\ref{SEC:VGR}): rendered node-link diagrams serve as input to VLMs for structural QA and knowledge-grounded queries. 
\textbf{Graph Learning} (\S\ref{sec:vgl}): visualizations provide features that either augment GNN representations or feed directly into vision encoders. 
\textbf{Reasoning and Learning for Scientific Graphs} (\S\ref{sec:sci_graph}): VLMs interpret molecular depictions for structured outputs, while visual encoders extract representations for property prediction.}
    \label{fig:pipeline}
\end{figure*}

This survey provides a first systematic overview of this emerging area, which we term \textit{vision meets graphs}. We introduce the Rendering-Perception-Inference (RPI) 
framework as 
a diagnostic lens, and organize 
work into three threads: \textbf{Vision for Graph Reasoning} (\S\ref{SEC:VGR}), \textbf{Vision for Graph Learning} (\S\ref{sec:vgl}), and \textbf{Scientific Graphs} (\S\ref{sec:sci_graph}). Fig.~\ref{fig:taxonomy} summarizes our taxonomy of the literature, while Fig.~\ref{fig:pipeline} illustrates representative input--model--output pipelines for each paradigm.
We conclude by discussing open challenges and outlining how this line of research can contribute to broader visions such as an AI social scientist~\cite{chen2025ai} that extracts and analyzes networks from scientific literature, and an AI scientist~\cite{ghafarollahi2025sciagents} that reasons with scientific graphs as human experts do.

\section{Background}
In this section, we introduce notation and review how graphs are represented for GNNs, for LLMs, and for vision models.
\subsection{Graphs and Their Representations}
A graph $\mathcal{G} = (\mathcal{V}, \mathcal{E})$ consists of a node set $\mathcal{V}$ with $|\mathcal{V}| = n$ and an edge set $\mathcal{E} \subseteq \mathcal{V} \times \mathcal{V}$.
Nodes may carry attributes $\mathbf{X} \in \mathbb{R}^{n \times d}$, and edges may also have features or weights.
For machine learning, this discrete structure must be converted into a representation suitable for computation.
GNNs operate directly on $\mathcal{G}$, learning node embeddings through neighborhood aggregation:
\begin{equation}
\mathbf{h}_v^{(l+1)} = \phi\left(\mathbf{h}_v^{(l)}, \bigoplus_{u \in \mathcal{N}(v)} \psi(\mathbf{h}_u^{(l)}, \mathbf{h}_v^{(l)}, \mathbf{e}_{uv})\right),
\end{equation}
where $\mathcal{N}(v)$ denotes the neighbors of node $v$, $\bigoplus$ is a permutation-invariant aggregation, and $\phi, \psi$ are learnable functions.
This message-passing paradigm respects topology but aggregates locally, which leads to well-known expressiveness limitations~\cite{zhang2024expressive}.

LLMs take a different approach, serializing $\mathcal{G}$ into a token sequence $\mathcal{S}$ by a serialization function $\mathcal{T}: \mathcal{G} \rightarrow \mathcal{S}$ for language-based reasoning.
The common formats include edge lists, adjacency descriptions, and structured templates~\cite{wang2023can,luo2024graphinstruct,xu2025graphomni}.
A third option, the focus of this survey, is to render $\mathcal{G}$
as an image $\mathbf{I}$ via a visualization function. This visual representation makes graph structure accessible to vision models, but introduces variability through layout and styling choices. We formalize this representation in the next subsection.

\subsection{Graph Visualization}\label{sec:vis}
Graph visualization maps abstract topology to a perceptible form.
We formalize this as a rendering function:
\begin{equation}
\mathcal{R}: (\mathcal{G}, \mathbf{P}, \boldsymbol{\theta}) \rightarrow \mathbf{I},
\end{equation}
where $\mathbf{P} \in \mathbb{R}^{n \times 2}$ denotes node positions, and $\boldsymbol{\theta}$ denotes visual encoding parameters such as colors, shapes, and line styles.
The output $\mathbf{I} \in \mathbb{R}^{H \times W \times C}$ is an image.
Different choices of $\mathbf{P}$ and $\boldsymbol{\theta}$ yield visually distinct depictions of the same graph, a property central to robustness probes.
\begin{figure}[h]
    \centering
    \includegraphics[width=0.9\linewidth]{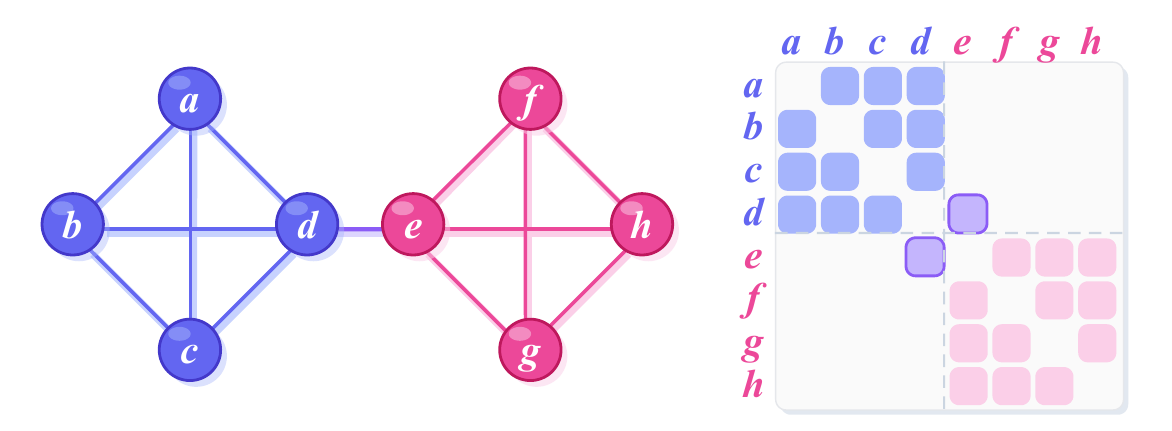}
    \caption{Two common graph visualization paradigms. Left: node-link diagrams. Right: a matrix view of the same graph.}
    \label{fig:graphvis}
\end{figure}

Two common visualization paradigms dominate practice (as shown in Fig.~\ref{fig:graphvis}).
\emph{Node-link diagrams} represent nodes as geometric primitives and edges as connecting lines.
Layout algorithms determine $\mathbf{P}$ by optimizing criteria such as edge length uniformity, node separation, and physical energy; common families include force-directed, spectral, and hierarchical methods~\cite{tamassia2013handbook}.
This paradigm offers intuitive correspondence to network structure but suffers from clutter in dense graphs.
\emph{Matrix views} arrange nodes along rows and columns, encoding edges as filled cells.
Reordering algorithms permute rows and columns to reveal block structure~\cite{behrisch2016matrix}.
Matrix views eliminate edge crossings but make path tracing less direct.

Scientific domains often adopt conventions that blend these paradigms.
Molecular skeletal diagrams follow node-link conventions with chemical semantics encoded through bond types.
Protein contact maps use the matrix paradigm, with cells indicating residue proximity.
These conventions, refined over decades, constrain $\boldsymbol{\theta}$ and simplify the visual vocabulary that models must interpret.

\noindent\textbf{RPI view: rendering, perception, and inference.}
In this survey, we use the RPI view as a diagnostic lens. \textbf{\textit{Rendering}} maps a discrete graph and visualization choices to images, $\mathbf{I}=\mathcal{R}(\mathcal{G},\mathbf{P},\boldsymbol{\theta})$, and determines which geometric and stylistic cues are available.
\textbf{\textit{Perception}} concerns establishing structural correspondences from pixels, such as identifying nodes, tracing edges, associating labels, and resolving endpoints under crossings, occlusion, and stylistic variation.
Perception does not require explicitly reconstructing the full adjacency matrix; many systems rely on partial or implicit structure as long as it supports the downstream objective.
\textbf{\textit{Inference}} produces task outputs from the perceived structure, ranging from \textit{predictive inference} for classification and regression to \textit{multi-step reasoning} for structural question answering and complex scientific reasoning.
Throughout this survey, inference serves as the umbrella term; we reserve reasoning for tasks requiring multi-step computation over graph structure. We use the RPI view to localize where methods intervene and where failures arise, without treating it as a taxonomy.

\section{Vision for Graph Reasoning}\label{SEC:VGR}

Graph reasoning is a nontrivial skill central to scientific research, requiring powerful structure perception and multi-step compositional reasoning capability. For foundation models, graph reasoning therefore serves as a particularly revealing testbed.  Typical graph reasoning tasks include structural understanding (e.g., counting, connectivity), algorithmic reasoning (e.g., shortest path, topological ordering), combinatorial problems, and multi-step reasoning over social-science graph \cite{wang2023can,li2024visiongraph,xu2025graphomni,wei2024gita,ai2024advancement}. 

From a human-centered perspective, graphs are rarely understood as raw adjacency.
Instead, we translate graphs into external representations that are easier to perceive or manipulate, such as textual descriptions, symbolic forms (e.g., matrices), and visual depictions.
Early LLM-era work naturally began with text, serializing graphs into sequences for prompting or instruction tuning~\cite{wang2023can,ren2024survey}.
This line produced valuable benchmarks and methods, and also revealed some limitations: reasoning performance varies with serialization format~\cite{xu2025graphomni}, and sequential encoding does not guarantee invariance to graph symmetries such as node reindexing or edge reordering~\cite{herbst2025lost}.
Rendered graphs offer a different trade-off.
They can expose global structure through spatial cues, yet they introduce variability through layout and style.
In this section, we review the efforts that probe whether models can robustly extract usable structure from pixels and then perform reliable reasoning on top of it, first in controlled graph-theoretic settings (\S \ref{SEC:VGR1}) and then in real-world graphs (\S \ref{SEC:VGR2}) that additionally require semantic comprehension.

\subsection{Graph-Theoretic Tasks}\label{SEC:VGR1}

A primary line of work evaluates VLMs on classical graph-theoretic problems rendered as images. Beyond reporting accuracy, these benchmarks are insightful because they expose where performance is gated, whether the model fails to read the graph structure from images, or fails to carry out multi-step reasoning after structure is perceptible.

Early work showed that vision is a viable channel for graph reasoning. GraphTMI~\cite{das2024modality} found that multimodal inputs consistently improve performance over text-only baselines. This suggests that visual renderings make certain structural cues easier to perceive than text serialization, even when the reasoning mechanism is otherwise similar. At the same time, the remaining gap indicates that simply adding an image does not remove the need for reliable graph perception and reasoning.
VisionGraph~\cite{li2024visiongraph} pushed further, introducing a benchmark spanning eight graph-theoretic task types, including connectivity, shortest path, and cycle detection, and it also includes basic topology perception probes for node and edge recognition.
It shows that the dominant failure mode in the zero-shot setting is structure perception: models frequently miss nodes or misbind edge endpoints, while supervised fine-tuning substantially improves both node and edge recognition. This indicates that performance is bounded by perceptual grounding, since downstream multi-step reasoning becomes much more reliable once the perceived structure is stable.
To reduce error cascades, VisionGraph also proposes Description-Program-Reasoning (DPR), an agent pipeline that first produces an explicit structural description and then executes algorithm-aware code for multi-step reasoning.
By externalizing intermediate structure, DPR makes downstream reasoning more controllable and reduces error cascades.
GITA~\cite{wei2024gita} scales evaluation with a larger dataset for vision-language graph reasoning.
A central observation is that combining visual and textual modalities outperforms either alone, with each excelling at different task types.
This complementarity is informative: visual inputs tend to stabilize global topology cues, while text can help with label semantics and instruction following.
GITA also proposes rendering-level augmentation, varying layout, node shape, and edge thickness, which improves generalization by expanding the rendering distribution seen during training.

VGCure~\cite{zhu2025benchmarking} probes deeper into complex graph reasoning, broadening evaluation to 22 tasks spanning node, edge, and graph levels.
Results confirm that accuracy degrades sharply as queries require more precise relational binding, such as the nested relation query and relation analogy query.
This highlights a coupling effect: small perception errors, such as a single missed endpoint, can invalidate an entire multi-step reasoning trace.
To address this, VGCure proposes MCDGraph, a structure-aware fine-tuning framework using masked graph infilling, contrastive graph discrimination, and graph description generation.
These objectives strengthen structure reading without relying on task-specific labels, which is especially relevant when new tasks are introduced but the perception problem remains the same. GraphVerse~\cite{sun2026graphverse} further broadens visual graph reasoning beyond single-image and answer-only evaluation, jointly assessing perception and reasoning under both single- and paired-image settings. It also introduces graph-centric image editing for controlled visual variations and a process-sensitive metric for reasoning quality.

Several additional benchmarks probe robustness and invariance. VisGraphVar~\cite{sartori2024visgraphvar} systematically varies layouts, styles, and imperfections such as node overlap, showing that accuracy is highly sensitive to such perturbations.
This indicates that many methods implicitly assume a narrow rendering distribution, and improvements may partly reflect overfitting to a particular visualization style.
Visual Graph Arena (VGA)~\cite{babaiee2025visual} tests invariance by asking whether models can recognize isomorphic graphs across different visualizations.
The gap between human and model performance suggests that learning layout-invariant structure from pixels remains challenging.
The VNA Benchmark~\cite{williams2024multimodal} evaluates basic network analysis tasks, including maximum degree identification, triad balance, and component counting.
These studies suggest that current models still struggle with reasoning. However, all three benchmarks use relatively sparse and simple visual styling. When visualization quality is limited, models may fail at perception before reasoning even begins.

\subsection{Real-World Graphs}\label{SEC:VGR2}
Beyond graph-theoretic reasoning (\S\ref{SEC:VGR1}), real-world graphs couple domain semantics with graph visualization.
We discuss two families: \emph{social-science graphs} and \emph{general-purpose visual graphs} such as flowcharts.

\paragraph{Social-science graphs.}
Social science relies heavily on graph visualizations: citation and collaboration networks reveal the structure of scientific communities, and system maps such as causal loop diagrams encode causal hypotheses in policy research. Social scientists extract insights from these visualizations.
However, some graphs exist as structured data but are communicated through visualizations, while others appear only as figures in papers and reports and must be extracted from images before analysis.
For the first case, several benchmarks in \S\ref{SEC:VGR1} evaluate whether VLMs can analyze social-science graphs as researchers do.
GITA evaluates on real-world social-science graphs, including PolBlogs graph and citation graphs; GraphTMI tests on citation networks; VNA uses synthetic graphs to test network analysis primitives commonly used in social science, including triad balance and component counting. Notably, GITA reports large improvements from visual inputs on PolBlogs, suggesting that vision may be particularly useful for social-science graph analysis.
For the second case, \cite{white2025using} evaluates whether VLMs can extract node and edge lists from system-map images (e.g., Causal loop diagrams, fuzzy cognitive maps) used in policy research, finding that node identification is relatively reliable but edge recovery (especially direction and polarity) remains a bottleneck.
Social-science graph reasoning has received less attention than graph-theoretic tasks, but we believe it deserves more. There are many potential applications; for example, an AI social scientist that automates science-of-science research would need to analyze citation and collaboration networks~\cite{chen2025ai}.

\paragraph{General-purpose visual graphs.}
Flowcharts, mind maps, route maps, and organizational charts convey rich semantic and visual cues through their rendering conventions, but their topologies are typically simple.
This shifts the bottleneck toward perception: Optical Character Recognition (OCR), symbol recognition, and binding text to the correct nodes and connectors.
\cite{ai2024advancement} constructs instruction-following data from web-crawled images across multiple graph types, showing that perception-based supervision improves downstream reasoning.
Graph-to-Vision~\cite{li2025graph} evaluates multi-graph joint reasoning and separates parsing quality from reasoning consistency, helping localize whether failures arise from visual graph reading, semantic interpretation, or subsequent inference. FlowGen~\cite{shi2026flowgen} complements these efforts with controllably synthesized flowcharts, allowing graph complexity and rendering style to be varied systematically. Its results show that both structural complexity and visual diversity remain challenging for current Multimodal large language models (MLLMs), while synthetic training improves flowchart parsing and question answering.

\subsection{Challenges and Future Directions}
\noindent\textbf{Rendering: controllable rendering distributions.}
Rendered graphs offer structural cues that are difficult to preserve under text serialization, but they also introduce a distribution over layouts and styles. From this perspective, rendering can be viewed as a computational interface between discrete graph structure and vision models, rather than merely a presentation choice.
Many benchmarks adopt simplified visualization schemes to isolate topology, which is useful for diagnosis, yet it narrows the rendering distribution and can limit how far conclusions generalize to tool-generated or human-facing diagrams.
Moreover, minimal renderings can inadvertently make perception harder by weakening endpoint cues, reducing node edge contrast, or removing textual anchors, so failures may mix up perceptual difficulty with downstream reasoning limits.
A promising direction is to build benchmarks with controllable rendering distributions spanning both perception-friendly and stress-test regimes, and reporting rendering parameters to allow comparison across studies.
Visualization research also offers useful tools here, including layout objectives, transparency, and emphasis techniques that reduce ambiguity in dense regions~\cite{tamassia2013handbook,meng2025seeing}.

\noindent\textbf{Perception: robust structure reading under ambiguity.}
Many failures labeled as reasoning errors originate earlier, during structure reading.
Models must localize nodes, follow edges, resolve endpoints at crossings, and integrate text labels, often under overlap and stylistic variation.
These challenges mirror spatial reasoning problems in natural images, where models must establish object correspondences under occlusion and depth ambiguity.
Progress in that domain shows that training on large-scale synthetic spatial VQA tasks can substantially improve spatial understanding~\cite{chen2024spatialvlm}; designing analogous synthetic tasks for graph-specific ambiguities, such as endpoint disambiguation and crossing resolution, is a promising direction.

\noindent\textbf{Inference: agentic visual graph reasoning and thinking with images.}
Beyond perception, the key challenge is to make multi-step inference controllable and checkable. A natural strategy is to expose intermediate states and decisions, including the graph structure being read, the computation being executed, and the evidence being selected.
Agent-based pipelines that separate structure extraction from algorithm execution offer one path forward~\cite{li2024visiongraph}, while GraphVista further explores an agentic VLM framework that uses a planning agent over hierarchical GraphRAG to route each query to textual descriptions, visual subgraphs, or both~\cite{han2025see}.
Broader visual programming systems show how code generation can orchestrate perception modules~\cite{suris2023vipergpt}.
When reasoning spans multiple graphs from heterogeneous sources (e.g., system maps alongside social networks), selectively routing each graph to a suitable perception encoder~\cite{zhang2025scope} may help.
A complementary direction, explored in recent work on visual reasoning~\cite{hu2024visual,qin2025chain,su2025thinking}, is to treat images as manipulable intermediate states rather than static inputs.
For graphs, this suggests operations like highlighting visited nodes, zooming into ambiguous regions, and local re-rendering to support multi-step reasoning while keeping each step inspectable. 
These capabilities point toward an \textbf{\textit{AI social scientist}} that uses visual intermediate states to extract and verify network structure from figures, and integrates with external analysis tools.

\section{Vision for Graph Learning}\label{sec:vgl}
In the previous section, we focused on VLM for graph reasoning, where a vision encoder and a language model work together. For many core graph tasks, however, practitioners still rely on representation learning with numeric outputs rather than text generation. Problems such as link prediction, node classification, and graph classification are usually solved by encoders followed by simple predictors~\cite{wang2019deep,lin2022spectral,jian2024rethinking,stoll2025graphbench}. This raises a basic question: can a vision encoder learn useful graph representations from rendered images? This section reviews early work that treats graph visualizations as inputs for representation learning.

GNNs have been known to have expressiveness limitations~\cite{zhang2024expressive}. For example, they are bounded by the Weisfeiler–Lehman test~\cite{wang2023empirical} and cannot detect properties such as graph biconnectivity that simple algorithms can compute~\cite{zhang2023rethinking}.
These limitations have often been visualized in past work to present to readers: "Look, these two graphs are not the same, but our advanced GNNs cannot distinguish them."
This observation suggests that graph visualizations encode certain information that message passing fails to capture, motivating the use of visual representations for graph learning. Here, the dominant objective is predictive inference, and perception is often implicit and optimized end-to-end.
A model may never emit an explicit intermediate graph, yet it still must establish correspondences between pixels and the underlying structural regularities that support prediction.
This is why rendering choices, such as the layout distribution and visual encoding, can matter even when the training objective is standard classification or regression.

\noindent \textbf{Vision-Enhanced GNNs.} Rather than replacing GNNs entirely, one line of work incorporates visual priors from graph visualizations to compensate for GNN limitations. GVN~\cite{wei2025open} renders subgraph visualizations centered on target edges and extracts features using a pretrained ResNet encoder. Analysis shows these visual features help distinguish links that 1-WL-bounded GNNs cannot. DEL~\cite{zhao2025graph} takes a different approach. It models graph visualizations as samples from a Boltzmann distribution, generates multiple visualizations, and extracts hand-crafted features from them to feed into GNNs. Hand-crafted features can be seen as the simplest form of visual encoder, which reduces ambiguity but discards much information. Even so, these features provide theoretical advantages, helping distinguish non-isomorphic graphs that 1-WL-bounded GNNs cannot. Both methods suggest that graph visualizations carry structural signals that complement message passing, though neither offers a purely visual solution.

\noindent \textbf{Vision Encoders for Graph-Level Tasks.} A more direct approach uses vision encoders to process graph visualizations as standalone inputs. Recent work~\cite{zhao2025underappreciated} shows that on graph-level tasks requiring global structure understanding, vision encoders pretrained on natural images can outperform GNNs. This suggests that spatial arrangements from visualization algorithms encode patterns that local message passing cannot capture. It also shows that vision encoders can serve as graph encoders, which could enable their use as backbones in graph-specialized VLMs.
However, current evidence remains limited to graph-level tasks. Whether vision-based approaches can match or exceed GNNs on node-level and edge-level tasks, where local structure matters more, remains an open question. Furthermore, all existing methods rely on vision encoders pretrained on natural images, which lack inductive biases for graph-specific patterns such as centrality, motifs, or community structure.

These methods have been evaluated on datasets spanning citation networks, social networks, and biological graphs, demonstrating applicability across both social science and natural science domains.

\subsection{Challenges and Future Directions}

\noindent\textbf{Rendering: theoretical understanding of what survives into pixels.}
Graph visualization is a structured but lossy channel. Different graphs can map to visually similar images, and the same graph can look different under different layouts and styles. 
For graph learning, an important theoretical question is which structural properties are preserved, amplified, or erased by a rendering distribution, and when those properties are learnable by standard vision encoders.
Progress here would turn rendering from an implementation detail into a design variable, enabling principled choices of layouts, multi-view renderings, or task-aware depiction rules that improve expressiveness.

\noindent\textbf{Perception: graph-native visual pretraining and invariance.}
Existing results are largely based on vision encoders pretrained on natural images, yet graph visualizations are symbolic and rule-governed.
A promising direction is graph-native visual pretraining that explicitly targets invariance central to graph learning, such as robustness to node reindexing, layout changes, and cosmetic style shifts, while learning sensitivity to true structural differences such as motifs and global connectivity patterns.
Well-chosen pretraining objectives can also clarify what the encoder is actually reading, reducing the risk that gains come from dataset-specific rendering artifacts.

\noindent\textbf{Inference: moving beyond graph-level prediction and isolating when vision helps.}
Evidence for purely visual encoders is powerful on graph-level tasks that reward global pattern recognition; it remains unclear whether similar benefits extend to node-level and edge-level prediction, where locality, semantics, and calibration matter.
The next step is to identify which task families and graph characteristics benefit most from visual cues, and whether hybrids provide new information or mainly act as regularizers. Such studies would also inform evaluation design, separating improvements due to better structure reading from improvements due to downstream prediction.

\section{Reasoning and Learning for Scientific Graphs}\label{sec:sci_graph}
Scientific domains present a distinctive setting for vision-based graph reasoning and learning.
Molecular diagrams, protein contact maps, and reaction schemes all follow standardized conventions refined over decades of scientific practice.
These conventions constrain both layout and visual encoding (the $\mathbf{P}$ and $\boldsymbol{\theta}$ of \S \ref{sec:vis}), resulting in near-canonical structure-to-pixel mappings that encode domain knowledge while dramatically reducing perceptual ambiguity.
The same visual vocabulary that enables expert reasoning thus becomes accessible to visual models, opening a path toward foundation models that interpret scientific structure as scientists do.

Two domains have received the most attention: molecular structures and proteins.
Molecular depictions follow node-link diagrams with atoms as nodes and bonds as edges, arranged according to well-established chemical drawing rules that chemists use daily in papers, textbooks, and laboratory notebooks.
Protein structures, by contrast, are often represented either as rendered 3D structure images or as contact and distance maps, which are matrix views where entry $(i,j)$ encodes spatial proximity between residues $i$ and $j$.
These two domains thus exemplify both major visualization paradigms introduced in \S \ref{sec:vis}, providing natural testbeds for vision-based methods.
This section examines how these conventions support both reasoning over rendered structures (\S\ref{sec:sci_reasoning}) and representation learning from visual inputs (\S\ref{sec:sci_learning}).

\subsection{Vision for Scientific Graph Reasoning}
\label{sec:sci_reasoning}

VLMs and MLLMs have opened new possibilities for scientific graph reasoning. These models can directly interpret visual depictions of scientific graphs and generate natural language responses, from describing properties to extracting symbolic representations. Early work focuses on perception and question answering, but the broader goal is to enable reasoning over scientific structures, such as predicting properties or understanding reactions.
This subsection surveys such reasoning approaches.

The molecular domain has seen the most activity, with research spanning tasks from focused applications to ambitious unifying frameworks.
A single molecule admits three coupled representations: a molecular graph $\mathcal{G}$, a symbolic string form such as SMILES, and a 2D depiction image $\mathbf{I}$ rendered under long-standing conventions.
The ability to map between these representations is the perceptual foundation for scientific graph reasoning.

Early efforts primarily target perception: aligning rendered graphs with symbolic forms. GIT-Mol aligns image, graph, and text modalities for molecule captioning and image-to-SMILES conversion~\cite{liu2024git}, while Mol2Lang-VLM focuses on generating natural language descriptions from molecular images~\cite{tran2024mol2lang}.
More recent work moves beyond perception to multi-step reasoning. ChemVLM~\cite{li2025chemvlm} trains on a bilingual multimodal dataset of molecular structures, reactions, and chemistry examinations, with evaluation suites spanning structure recognition, multi-step reasoning, and property prediction.
Strong performance across these tasks, substantially outperforming general-purpose VLMs, demonstrates that domain-focused training can yield models proficient in both perception and chemical reasoning.
ChemDFM-X~\cite{zhao2024chemdfm} extends the notion of rendering beyond images to include five chemical modalities (2D graphs, 3D conformations, images, mass spectra, and infrared spectra), and demonstrates cross-modal collaboration where spectra provide structural hints while symbolic representations constrain invalid options.
These results lend support to the premise that visual conventions encoding chemical knowledge are learnable, offering concrete progress toward foundation models that interpret scientific structure as scientists do.

As work in this area expands, the efficiency and evaluation have drawn more attention.
TinyChemVL observes that molecular images contain large uninformative backgrounds and proposes visual token reduction to improve efficiency, alongside reaction-level benchmarks that test more complex reasoning~\cite{zhao2025tinychemvl}.
MolVision provides a systematic evaluation across multiple VLMs, finding that visual information alone is insufficient: competitive performance requires multimodal fusion with textual context~\cite{adak2025molvision}.

Protein structure reasoning presents a contrasting picture.
While molecular VLMs and MLLMs have grown rapidly, only a handful of studies have explored whether similar approaches transfer to proteins.
Recent efforts include Geneverse, which finetunes LLaVA on protein structure images rendered from AlphaFold predictions for protein function inference~\cite{liu2024geneverse}. LiveProteinBench evaluates whether adding multi-view structural images improves LLM performance on protein property and function prediction, and reports that structural inputs often fail to help and can even degrade performance relative to sequence-only baselines~\cite{rong2025liveproteinbench}.
This negative result is informative rather than discouraging.
One plausible explanation is a perception and alignment gap: if the model is not trained to reliably read structural cues from these renderings, adding images can act as nuisance variation and burden cross-modal fusion, especially when the language or sequence channel already provides strong signals.
A second possibility is a rendering adequacy gap: common 2D views may not preserve the functionally relevant 3D information required by the inference target.
Disentangling these failure modes remains an open challenge and motivates depiction-aware training and evaluation that explicitly separates depiction reading, fusion quality, and downstream scientific inference. 

Visual reasoning also extends to materials science. Scanning transmission electron microscopy (STEM) images can be viewed as physically grounded renderings whose appearance is governed by an instrument-specific image formation process. MicroscopyGPT~\cite{choudhary2025microscopygpt} shows that VLMs can predict crystallographic descriptors from such renderings, including lattice parameters, space-group information, and atomic coordinates, providing early evidence of perceptually grounded reasoning and suggesting that more complex multi-step inference may follow.

\subsection{Vision for Scientific Graph Learning}\label{sec:sci_learning}

Besides reasoning, another line of work learns visual representations of scientific graphs for prediction tasks.
Rather than generating language, these methods train encoders to produce embeddings that capture structural and chemical information, which are then fed to task-specific predictors.

For molecular graphs, early work applies visual encodes directly to 2D molecular images for property prediction~\cite{goh2017chemception}.
ImageMol~\cite{zeng2022accurate} improves substantially by pretraining on millions of unlabeled molecular images with chemically motivated objectives such as structure clustering and mask-based contrastive learning. More recently, MolEmb~\cite{zhao2026molemb} explores a different route by adapting multimodal large language models into reusable molecular embedding models. It combines molecular depictions with symbolic representations and allows natural-language context to condition the resulting embeddings, extending visual molecular representation learning toward multimodal and context-aware representations.
Beyond single-molecule prediction, visual representations naturally extend to relational tasks involving multiple molecules. For drug-drug interaction prediction, DDVR-DDI fuses two drug depictions into a single image to capture potential interaction interfaces~\cite{xie2025predicting}, while S$^2$VM blends local visual fragments from drug pairs to learn interaction-relevant embeddings~\cite{ma2025selfsupervised}.

For proteins, contact and distance maps encode pairwise residue proximities as matrices, providing a complementary view that avoids the edge crossings and node occlusion common in node-link diagrams.
CNNs have been applied to such matrix representations for sequence profile prediction~\cite{chen2019improve} and residue-level architecture segmentation~\cite{eguchi2020multi}.
Compared to molecules, however, visual representation learning for proteins remains underexplored.
This asymmetry may help explain the reasoning failures reported in \S \ref{sec:sci_reasoning}: without first establishing that visual encoders can reliably extract structural cues from protein renderings, multimodal reasoning has no solid perceptual foundation to build on.
Protein-specific visual pretraining, as ImageMol provided for molecules, may be necessary before VLMs and MLLMs can effectively understand and reason over protein structures.

\subsection{Challenges and Future Directions}\label{sec:sci_challenge}

\noindent\textbf{Rendering: scientific graph visualization as foundational infrastructure.}
Scientific graphs are produced by established tools (e.g., RDKit\footnote{\url{https://www.rdkit.org}}, PyMOL\footnote{\url{https://www.pymol.org}}) following conventions that evolved primarily for human readers.
Yet the design and iteration of these tools has rarely been treated as a research problem in its own right.
As rendering is the first stage of the pipeline, limitations here propagate downstream: a poorly rendered structure leaves perception and inference with no solid ground to build on.
Systematic benchmarks that evaluate how rendering choices affect model comprehension would help identify bottlenecks and guide tool improvements.
Such infrastructure is a prerequisite for any AI scientist who must reliably read scientific figures at scale.

\noindent\textbf{Perception: domain-specific visual pretraining.}
\S\ref{SEC:VGR} and \S\ref{sec:vgl} identified perception as a recurring bottleneck: models must reliably read graph structure before reasoning or learning can proceed.
Scientific graphs offer a partial advantage, as standardized depiction conventions reduce layout and style variation.
Yet domain-specific visual pretraining remains essential.
Molecules benefit from large-scale efforts like ImageMol, while proteins lack an equivalent foundation, which may explain the negative results in \S\ref{sec:sci_reasoning}.
Many other scientific domains with standardized conventions, such as metabolic pathways and gene regulatory networks, remain unexplored and would benefit from similar investments.

\noindent\textbf{Inference: from exam-style QA to scientific workflows.}
Current evaluations often emphasize exam-style question answering, but practical scientific workflows demand more (e.g., synthesis planning, pathway analysis, or materials property prediction): models must produce extracted structures that respect domain constraints.
These constraints are graph-structured (e.g., valence rules, regulatory directionality, or crystal symmetry) and should be checked by external tools (e.g., rule checkers or simulators), not just language reasoning.
Achieving the broader vision of an AI scientist calls for benchmarks that require models to output explicit graph structures, invoke domain tools for validation, and trace errors back to rendering or perception when results fail.

\section{Conclusion and Outlook}
This survey has offered the first unified treatment of \emph{vision meets graphs}, introducing a diagnostic framework organized around rendering, perception, and inference. Across all three threads, perception emerges as a recurring bottleneck: models must reliably extract graph structure from pixels before reasoning or learning can proceed.
Looking ahead, we see opportunities in graph-native visual pretraining, active visual reasoning where models manipulate rendered graphs during inference, and tighter integration with domain tools for structure verification. Scientific domains offer promising testbeds, as standardized depiction conventions reduce perceptual ambiguity while practical workflows provide meaningful evaluation. Beyond social and natural science, engineering domains with established visual conventions remain open for exploration. These directions point toward broader visions such as an AI social scientist and an AI scientist that read and reason over graphs as humans do.
\section*{Acknowledgements}
This work was supported in part by the National Natural Science Foundation of China (grant No. 92470113).

\bibliographystyle{named}
\bibliography{ijcai26}

\end{document}